# Image Guidance for Robot-Assisted Ankle Fracture Repair

Anthony Wu, Jayaram Mandavilli, Asef Islam

Mentors: Dr. Jeff Siewerdsen and Dr. Wojtek Zbijewski

May 9, 2020

# 1. Introduction

**Summary**

This project concerns developing and validating an image guidance framework for application to a robotic-assisted fibular reduction in ankle fracture surgery. The aim is to produce and demonstrate proper functioning of software for automatic determination of directions for fibular repositioning with the ultimate goal of application to a robotic reduction procedure that can reduce the time and complexity of the procedure as well as provide the benefits of reduced error in ideal final fibular position, improved syndesmosis restoration and reduced incidence of post-traumatic osteoarthritis. The focus of this product will be developing and testing the image guidance software, from the input of preoperative images through the steps of automated segmentation and registration until the output of a final transformation that can be used as instructions to a robot on how to reposition the fibula, but will not involve developing or implementing the hardware of the robot itself.

**Background**

Ankle fractures occur with a frequency of around 174 cases per 100,000 adults per year, with over 5 million yearly cases in the U.S. alone (Goost et al), affecting mainly young active people and the elderly. Ankle fractures most commonly involve a fracture in the lower fibula which can also result in disruption of the syndesmosis, or the alignment of other bones and ligaments within the ankle joint, if the fibula is displaced. This is due to the displacement of the lower fibula causing damage and forceful shifting of these ligaments and other connective tissue. This can lead to many long-term complications including post-traumatic osteoarthritis (PTOA) if the proper syndesmosis is not restored.

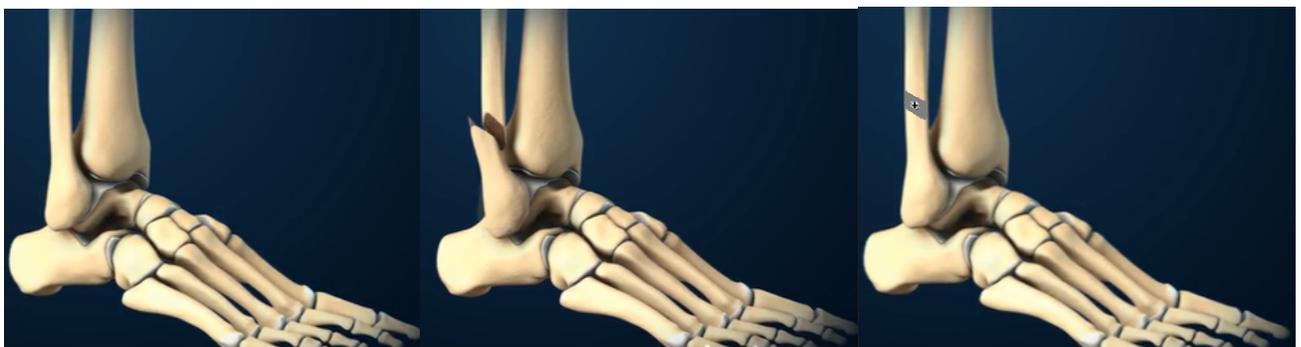

*Figure 1: Healthy ankle (left) and ankle after fracture (middle) with fibular fracture and displacement of the lower fibula in ankle joint, along with post-treatment ankle (right) with fixed fracture and restored fibula position.*

The proximal fracture can be treated with relatively little difficulty such as by fixation with metal screws. However, the second part of the surgery is a fibular reduction, or movement of the lower fibula back into proper position to reconstruct the joint and restore syndesmosis.

The current standard of care for this procedure is a visual estimation by the surgeon to determine where to place the fibula and screw-based fixation of the syndesmosis. In one study over 20% of patients who underwent reduction via this method were shown through CT scans shortly afterward to have syndesmotic malreduction, which was defined as a greater than 2 mm widening of syndesmosis compared to the patient's other, healthy ankle (Nagvi et al). The most common reason for re-operation in the weeks immediately following the initial ankle fracture surgery is syndesmotic malreduction (Ovaska et al). Furthermore, the incidence of PTOA in ankle fracture patients is as high as 70% (Mehta et al). Evidently, the standard of care for reduction is inadequate for proper syndesmotic restoration and prevention of PTOA.

## 2. Technical Approach

**Automated Segmentation**

The first step in an image-guided approach is fast, automated segmentation of the pre-operative images in order to identify and save the relevant anatomical features, in this case, the fibula and other bones of the ankle. A promising existing approach to automated segmentation is active shape models (ASMs) (Brehler et al). An ASM model is trained by taking in as input a training set of several segmentations of a single bone from different patients and then performing a PCA-based analysis to identify and characterize the principal modes of variation in the surface morphology of the bone within the population of training samples. The model comprises a set of basis functions for each principal component of variation, which can be tuned and create a deformable template that can be mapped onto any new example by minimizing residuals. Thus, this can be applied to automated segmentation by using this template object in order to map onto a new image and segment the bone given some initialization.

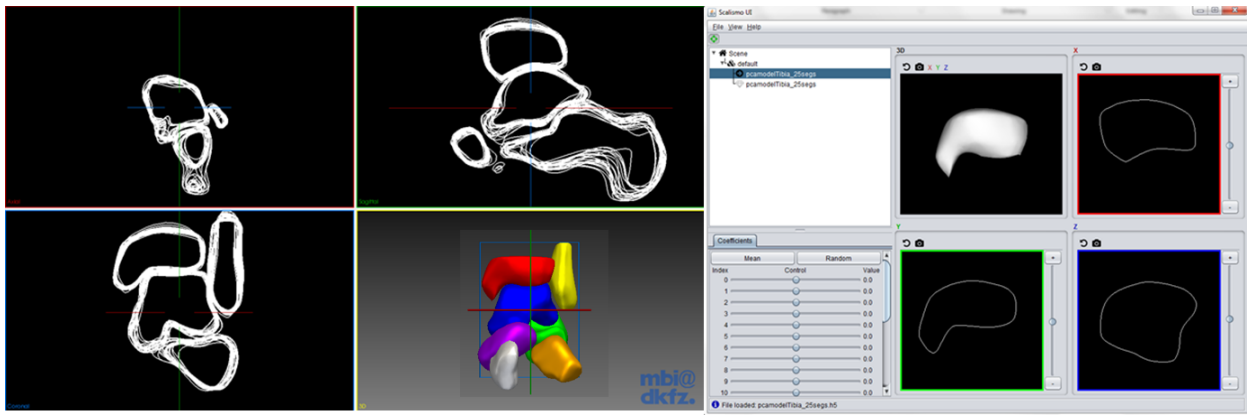

*Figure 2: Example of multiple registered 3D CT segmentations of ankle bones (left) which are used to construct ASM (right).*

After creating the segmentations, the positions of many corresponding landmark points (on the order of the minimum number of vertices on the segmentations on the training set) is also annotated on them.

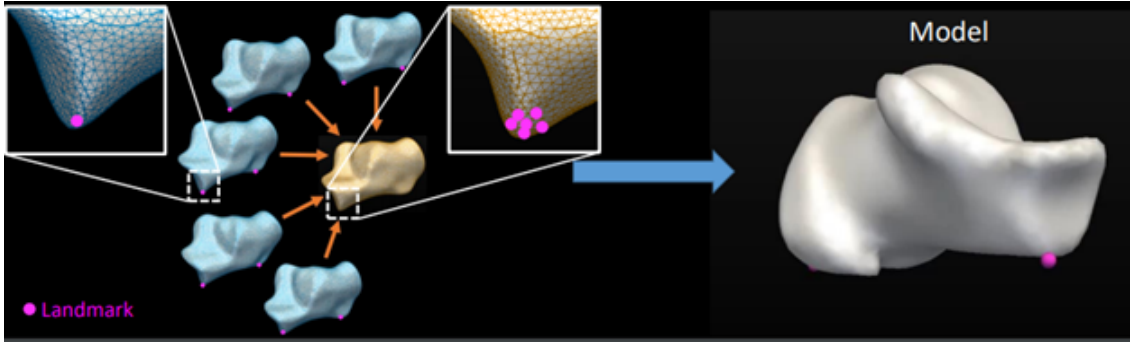

*Figure 3: Construction of ASM model from training segmentations and landmark points (Brehler et al).*

These points are represented in 3D space and are fed into the model to summarize the morphological variance in the training population. Each entire training segmentation can thus be represented as a point in 3n space, where n is the number of landmark points. Then, Principal Component Analysis (PCA), a common algorithm for variance-preserving dimensionality reduction, is used to project this representation into the t-dimensional subspace that preserves the most variance of the data. Hence, the dimensionality of the training data has been reduced from 3n to t principal modes of variance, where t is specified for the particular application. Any example x in the training set can then be represented as the mean plus the deviation from the mean, or $\bar{x}$ + Pb, where $\bar{x}$ is the average shape of the whole training set, P is the projection matrix from 3n to t produced by PCA, and b is a t-vector representing the deviation of the particular example from the mean along each mode. Thus, this can also be thought of as representing a deformable template that can be morphologically transformed by adjusting the variation from the mean along each mode in the b vector to encompass all of the shape variation of the bone within the population.

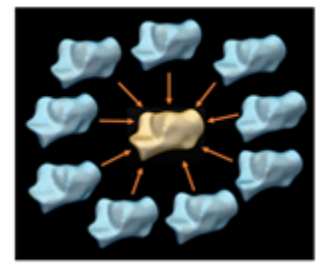

n landmark points
t principal components

$x \approx \bar{x} + Pb$
x and $\bar{x}$: 3n x 1
P: 3n x t
B: t x 1

The below examples show the mean shape, as well as the two shapes representing ± 10 standard deviations from the mean for the 7 bones of the ankle. Although some differences are notable, for many of these bones the shapes at 10 standard deviations appear more visually similar to the mean than one would expect for that many standard deviations. This reveals that in general the morphology of these bones is quite similar in the population and the differences are very subtle. The tibia and fibula are the two long shin bones and extend from the ankle joint up the leg all the way to the knee joint, but in this case, since only their surface at the ankle joint is of interest their segmentations are truncated at a set distance up from the ankle.

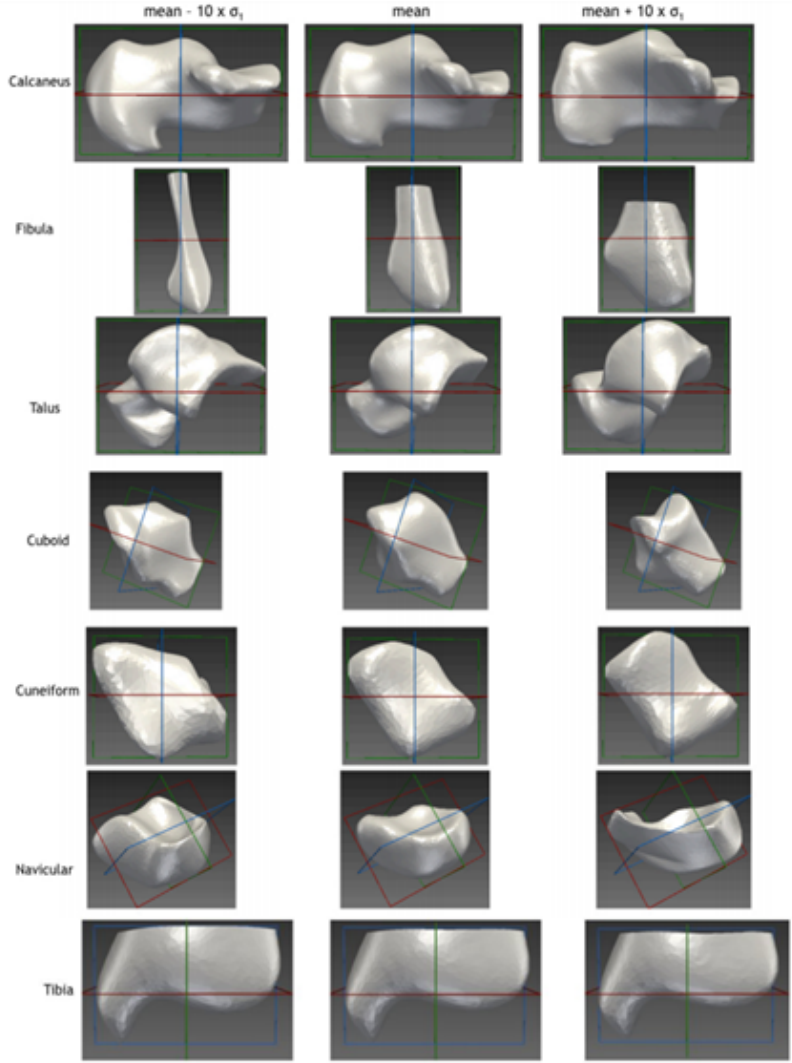

*Figure 4: Example shapes for 7 bones of the ankle.*

ASMs can be useful not only in understanding the variations between people in the shape of their bones, but more interestingly for surgical applications and particularly automatic segmentation. From a CT scan of a patient's ankle, the gradient version of the image can be taken to highlight just the edges, where the gradient is high. The mean models of each bone are then initialized to the image by rigid registration to the gradient edges. Then, in an iterative process to update the shape vector b, the ASM for each bone is deformed and shifted in the positions of the vertices so that they best fit the edges of the image.

ASMs can be improved even further by updating them to Coupled Active Shape Models (cASMs). Traditional ASMs are trained independently for individual bones and separately initialized, however, as a result, they are susceptible to errors in the narrow articular joint spaces between bones if there is poor contrast in the image and can result in producing segmentations with an overlap in adjacent bones. cASMS are trained with consideration of multiple bones simultaneously and incorporate the spatial relations and articular joint widths between adjacent bones as proximity constraints, and thus they are able to improve on the accuracy of segmentation within these joint spaces and prevent overlaps.

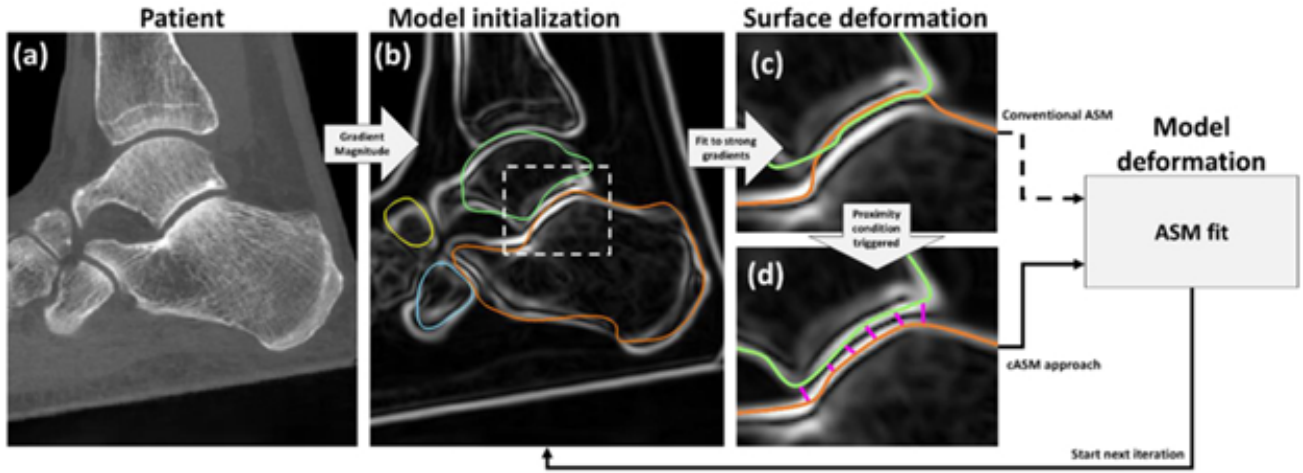

Figure 5: Outline showing main steps of cASM code: computation of gradient image, initialization by course registration to gradient image, and iterative surface deformation with proximity constraints. (credit: Brehler et al)

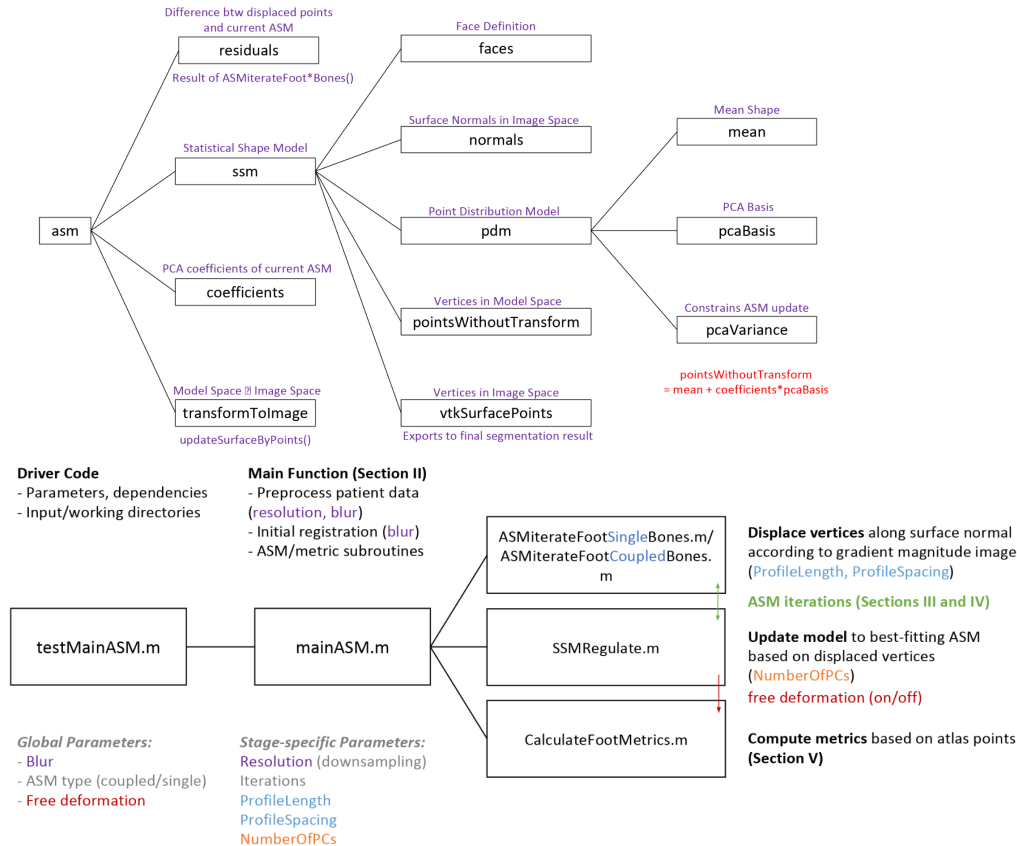

Figure 6: Diagram breaking down source code of cASM (credit: Wojtek Zbijewski, Qian Cao)

While in this work the cASMs were evaluated in high-resolution cone-beam CT (CBCT) images, they have not yet been tested in pre-operative C-arm images for 2D-3D reconstruction. Thus, an important step in adopting the cASM approach for this project will be to evaluate its accuracy in segmenting C-arm images and improving it to do so, including improving the initializations and adding corrections for noise and artifacts.

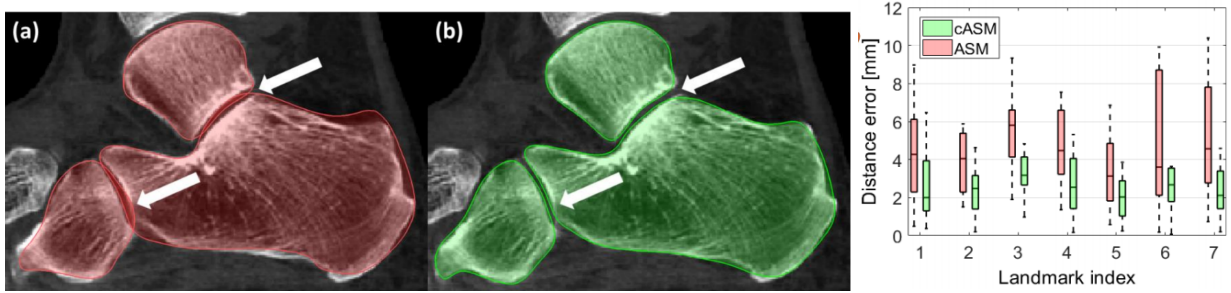

*Figure 7: Comparison of traditional ASMs (red) and cASMs (green) showing reduced overlap and error in segmentations within articular joint spaces as well as reduced error in landmark positions.*

A deep learning approach was also developed. The two deep learning ideas we looked into were creating a model to completely do the segmentation (for each bone) and creating a model that would initialize cASM (so cASM would work on the results of the network making cASM a layer in the neural network). We looked into the second idea and developed it a little bit but decided to pursue the first idea further starting off with the fibula. We ended up building multiple networks and adjusting the parameters to find the best accuracies.

Before training our models, a lot of data pre-processing was required. First, for each slice, we averaged 7 slices around it. This reduced noise in the training data theoretically resulting in a better model. Next, we applied a zoom in function to all of our slices. This shrunk each of the axes of all of our slices by a factor of 4. Although this reduced the quality of our images, this procedure also reduced the computational burden of training our model. This allowed us to make our model itself more complex without losing much information in the images.

Our first attempt with deep learning was to build a simple convolutional neural network (CNN). Our first idea was to create 3 models for each view (axial, sagittal, and coronal) and predict if each slice of each view either had a fibula or not. If the pixel in the ith axial slice, jth sagittal slice, and kth coronal slice all had a fibula, then we could say pixel i, j, k of the original image also belonged to the fibula.

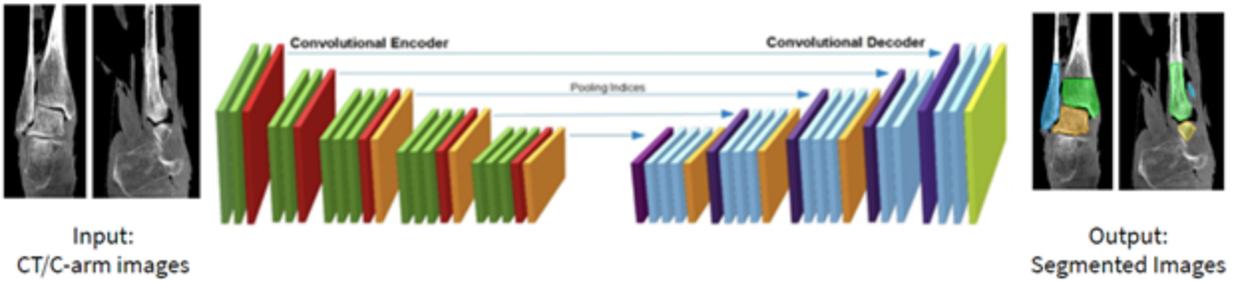

*Figure 8: This is the typical format for the CNN architecture.*

Our second approach was utilizing a U-net. This is a type of CNN that is very well documented. It is known to work well with limited data which is what we had, that is why we decided to pursue it. We used coronal 2D slices as the inputs to the network. Given that 0.3% of pixels in all C-arm images were labeled as a fibula pixel, a neural network could theoretically achieve 99.7% accuracy if it labeled every pixel as a non-fibula pixel. Hence, for our loss function, we used a cross entropy function, which penalizes a model significantly if it predicts a fibula pixel is a non-fibula pixel. With 200 epochs, we were able to achieve good initial results (as seen in the "Results" section below). The U-net seems to be the best architecture but we will be trying others in the future as well.

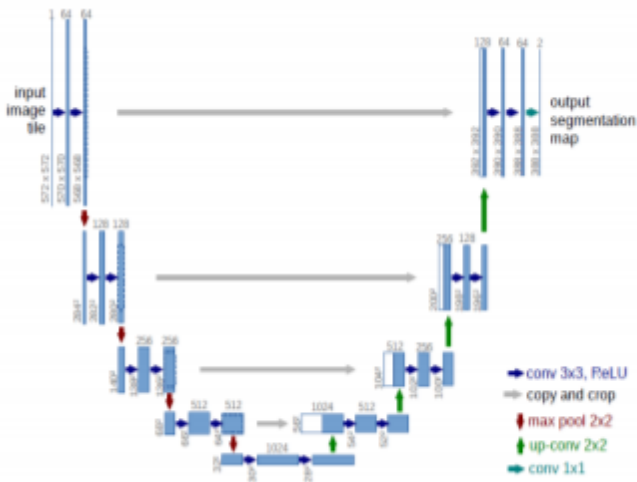

*Figure 9: This is the typical format for the U-net architecture.*

Other networks we will be looking into are the NAS-Unet, a deeper CNN, a Competitive Hopfield Neural Network (CHNN), etc. The CHNN uses a winner takes all approach and has been shown to work on medical images but may be too complicated and not much more beneficial than a U-net or NAS-Unet. Its applications were described by Cheng, Lin & Mao.

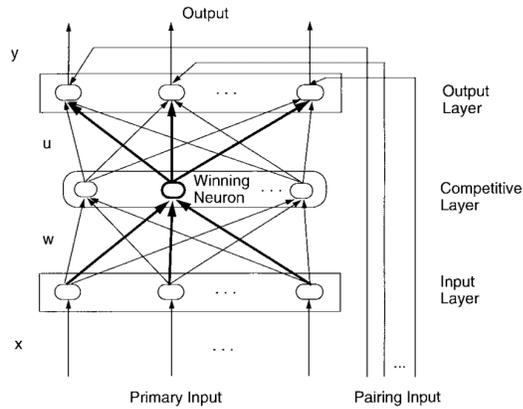

*Figure 10: This is the typical format for the CHNN architecture.*

The NAS-Unet was developed by Weng, Zhou, Li & Qui. They found that by using the U-net backbone they could create a more complicated architecture. They essentially replaced each block in the U-net with a unique cell architecture (one architecture for the downscaling layers and one for the upscaling layers but each cell would have unique primitive operations). This is where the name NAS-Unet comes from because NAS stands for neural architecture search. They used already developed machine learning algorithms to find the optimal cell architectures. Their results show that their networks achieved higher semantic segmentation accuracies than the classic U-net but the computational burden was much higher. The general NAS-Unet architecture can be seen below.

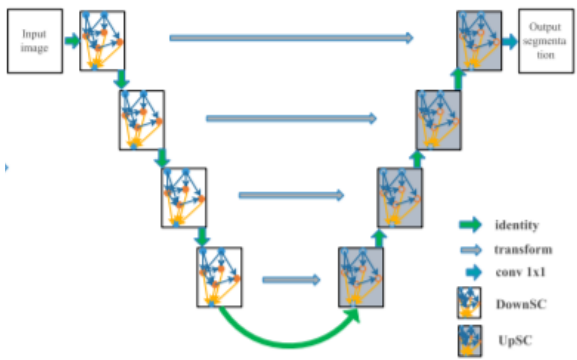

*Figure 11: This is the typical format for the CHNN architecture.*

In the future, we will either pursue these models separately or try to use the combined cASM-DL approach. We will also look into these other types of neural networks, especially the NAS-Unet. The model which yields the lowest error (measured by closest points to a mesh) for each of the bones will be used. We have code to obtain meshes for each segmentation that will allow us to calculate this error. It is possible that different approaches could be used for different bones (for example this could be as simple as changing some of the primitive functions or layers for each of the bones in their respective neural networks). After getting accurate segmentations, we will be able to proceed to the next stage.

**Registration**

After segmentation is completed, the instructions for reduction must be computed. This will be done by using the patient's healthy ankle as a guideline for how to reconstruct the injured ankle. The segmentation of the healthy fibula will be mirrored and flipped, and then the rigid registration from the injured to the healthy ankle will be computed. The resulting calculated transformation will be used as instructions for the robot on how to position the injured fibula.

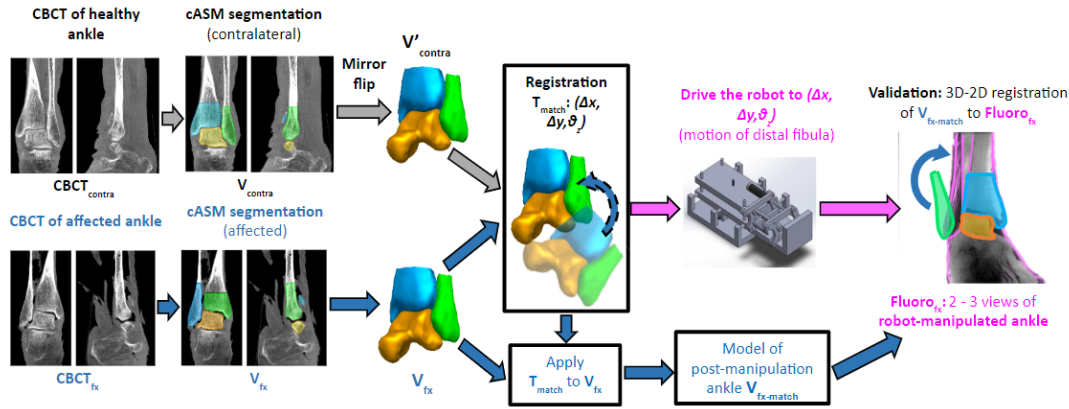

*Figure 12: Outline of approach (credit: Wojtek Zbijewski).*

## 3. Results

**Image Acquisition**
26 cadaver ankles were scanned with a Siemens Cios Spin mobile C-arm on 3/11/20 and classified into groups of high quality, low quality, and with metal artifacts as follows:

| High Quality: | Low Quality: | Artifacts: |
|---|---|---|
| 4_1 | 22_1 | 14_1 |
| 8_1 | 53_1 | 30_1 |
| 10_1 | 18_1 | 31_1 |
| 12_1 | 47_1 | 49_1 |
| 20_1 | 16_1 | |
| 24_1 | 37_1 | |
| 26_1 | | |
| 28_1 | | |
| 33_1 | | |
| 35_1 | | |
| 41_1 | | |
| 43_1 | | |
| 51_1 | | |
| 45_1 | | |
| 39_1 | | |
| 55_1 | | |

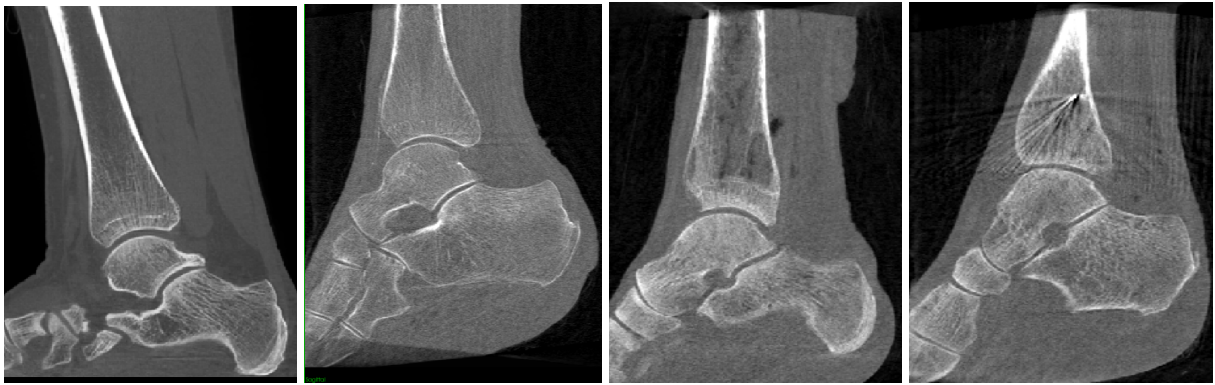

*Figure 13: Example of diagnostic-quality CBCT image (left) compared to C-arm images of high quality, low quality, and with metal artifacts respectively.*

This C-arm is designed for cost-effective intraoperative use, which is the envisioned application of this project, and thus the images are not of as high resolution or contrast as the diagnostic CBCT.

**Manual Segmentations**
In order to obtain ground truth segmentation, we utilized MITK workbench to segment all of our data. This allowed us to use these in our models as ground truth values.

**cASM**
The first step taken in running the cASM code was to run it on one of the same high-quality diagnostic CBCT images on which it was validated prior, just to ensure that we were able to run it properly and confirm that it works on these high-quality images. After some debugging of the code mainly to adapt it to the new local environment, the code was able to run properly.

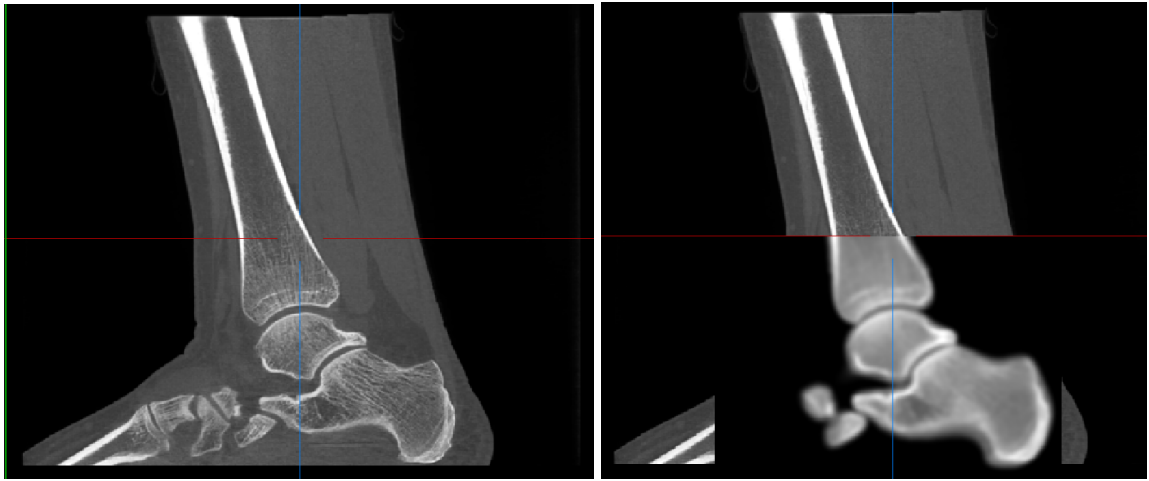

*Figure 14: Initial registration step to CBCT image showing target image (left) and overlaid alignment of mean image to target image (right).*

As can be seen, the initial registration of the mean image to the target image was highly accurate, and the mean image is almost perfectly aligned. Thus, not many iterations of surface deformation were needed to achieve a good fit.

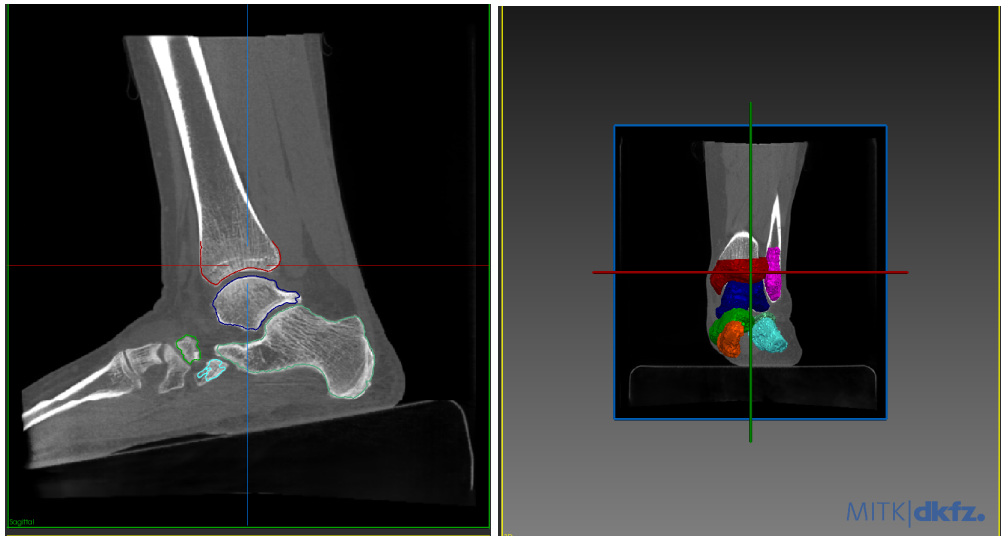

*Figure 15: Final segmentation result on CBCT image showing each bone in different color.*

As can be seen, the final segmentations for each bone are also highly accurate, with it being visually evident that the outlines of the segmentations correspond well to the true edges of the bones.

The attempt was then made to apply the cASM code to the C-arm images. However, the initial registration failed. As seen below, the coordinate frame of the C-arm data was completely different from that of the mean image, which was calibrated to the CBCT coordinate frame. Thus, it was necessary to first fix the initial alignment of the mean image with respect to the C-arm data to assist the initial registration.

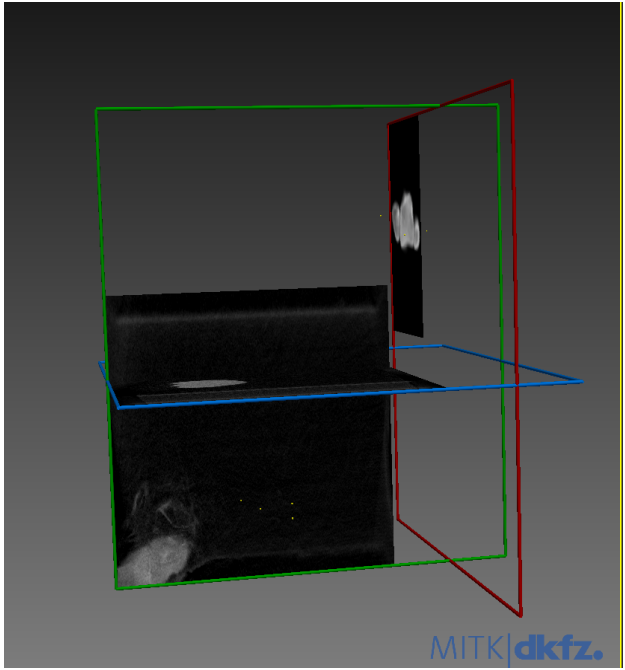

*Figure 16: This shows the differences between the C-arm and CBCT coordinates.*

The approach taken to better initialize the mean image to the C-arm coordinate system was to use a landmark point-based registration algorithm. At first a few landmark points were manually annotated and used, however later it was realized that a better approach was to use one of the manual segmentations of the bones and the corresponding mean shape as the point sets to provide a larger number of points and hopefully more accurate registration. In addition, at first a ICP algorithm was used in Matlab, however it was realized that ICP only supports rigid registration, which is without scaling. However, not only is the coordinate frame of the mean image different from the C-arm in terms of rotation and translation, but the voxel size is also different so a scaling component is necessary in the registration. Thus, a CPD algorithm was used instead. The initial CPD registration was decent but still needed some manual tweaking of both scaling and rotation and adjustment of the origin when saving as NiFTI format, as this affected the translation component.

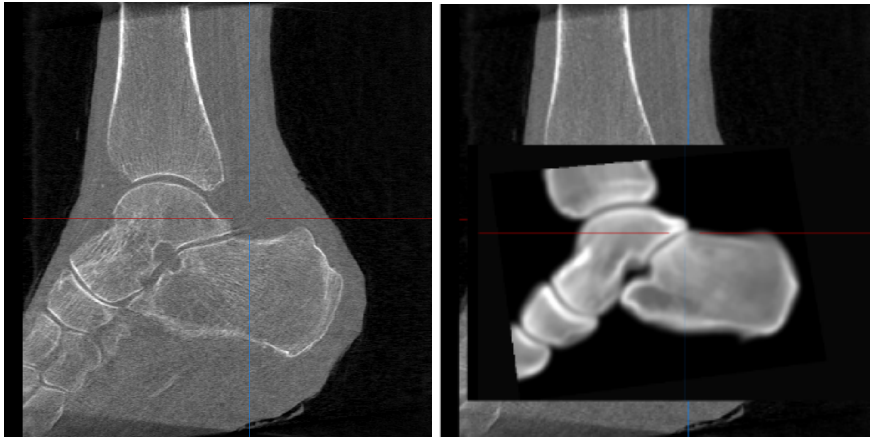
*Figure 17: Improved initial alignment of mean image to C-arm target image.*

The same transformation must also be applied to the meshes representing the mean shapes for each bone. At first it was attempted to simply apply the transformation matrix to the meshes in Matlab by computing the forward transformation to each vertex in the mesh. However, this resulted in a mesh with incoherent faces. Thus, the bones were instead re-segmented from the newly transformed mean image and the mean surfaces remeshed with the proper numbers of vertices.

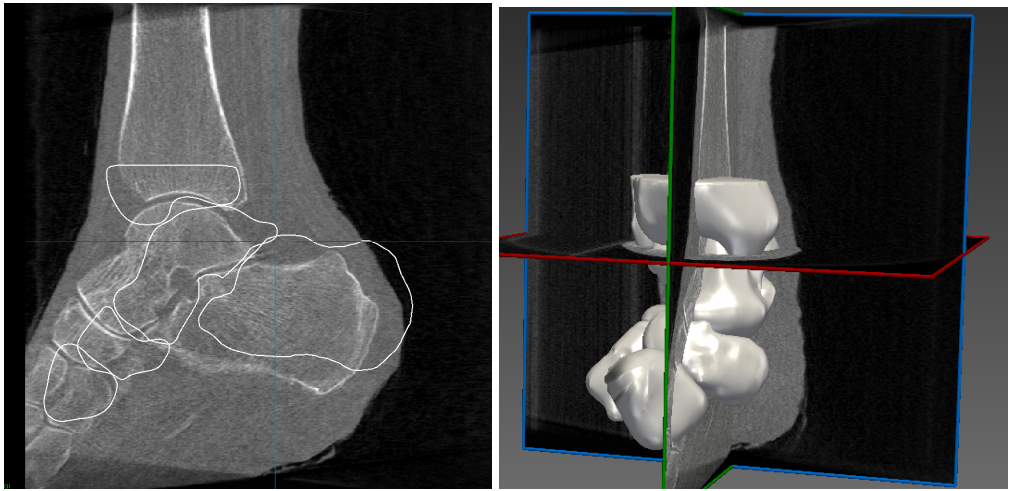
*Figure 18: Remeshed mean shapes overlaid on target C-arm image.*

cASM was then run on C-arm data with the newly transformed mean image and meshes. However, as of now it still does not run properly. A few problems persist: namely, in computing the normals to each vertex in the mean shapes during the constructions of the SSMs in order to know which direction to expand or collapse edges during surface deformation, some normals for a small number of points are returned as error values. It is unclear why this is occuring, but it may be attributable to some problems in the surfaces of the remeshed mean shapes. In addition, although the initial alignment of the mean shapes to the target image appears quite close, during the surface deformation process the mean shapes become again displaced out of the coordinate frame of the target image. These are problems that we are continuing to work on in order to achieve functionality of the cASM approach with C-arm images.

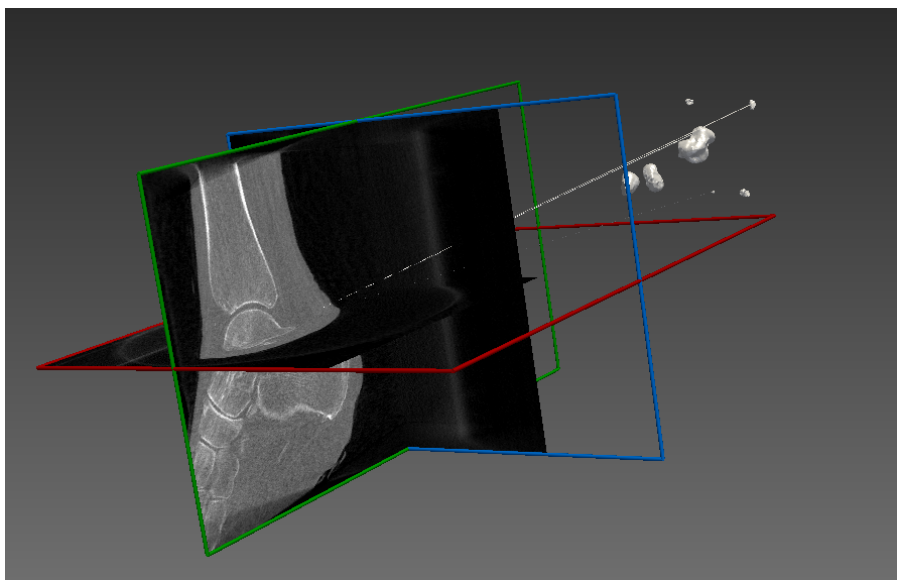

*Figure 19: It was attempted to find a registration between the C-arm and CBCT axes. However, as seen in the image, even after applying this registration the two did not align.*

**Neural Net**

**Simple CNN Approach**

After attempting the CNN approach described earlier, we could not get any image resembling a fibula. This low accuracy occurred because after 2-3 epochs, the loss and accuracy curves for training and test data diverged for all 3 slice views. This means that the models were overfitting too quickly which explains the very low final accuracies.

**Axial View:**

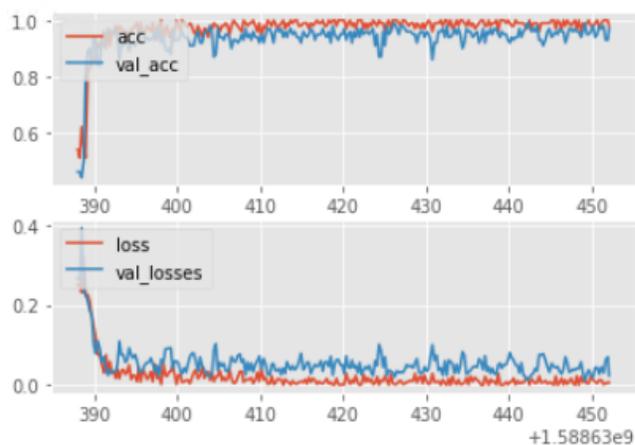

*Figure 20: These are plots of the accuracies and losses for both the train data (red) and test data (blue) with respect to epochs. This is for the model for the axial view slices.*

**Sagittal View:**

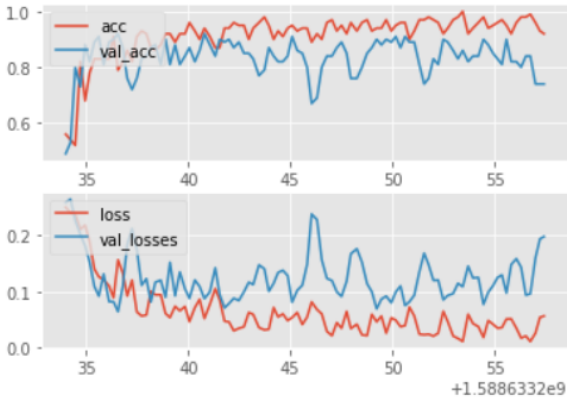

*Figure 21: These are plots of the accuracies and losses for both the train data (red) and test data (blue) with respect to epochs. This is for the model for the sagittal view slices.*

**Coronal View:**

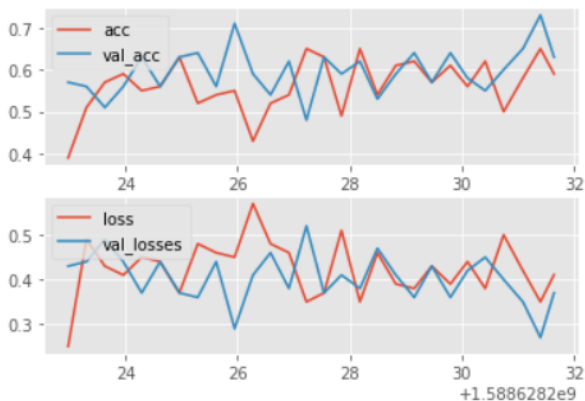

*Figure 22: These are plots of the accuracies and losses for both the train data (red) and test data (blue) with respect to epochs. This is for the model for the coronal view slices.*

**U-Net Approach:**

After the simple CNN approach failed, we implemented a U-net, training on 200 epochs. We used coronal view slices as inputs and obtained much better results.

**U-Net Segmentation (Ankle 24_1)**

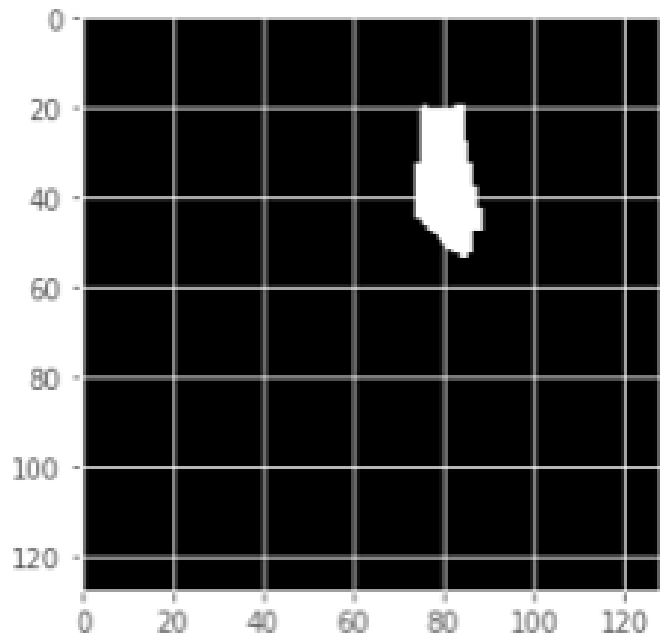

*Figure 23: This is an example of a fibula segmentation predicted by the U-net model. It is one of the coronal slices in sample 24_1.*

**Ground Truth (Ankle 24_1)**

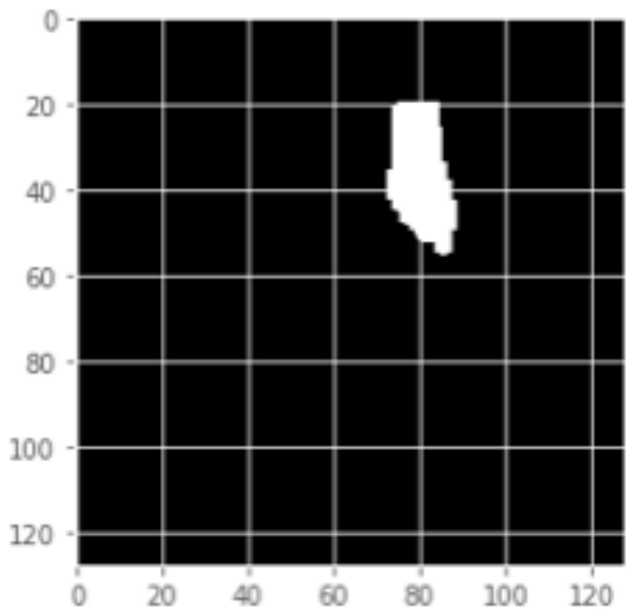

*Figure 24: This is the ground truth segmentation of the fibula obtained from our manual segmentations. It is the same slice in sample 24_1 as the image above. As seen in the images, there are some clear variations but the general shapes are very similar.*

**C-Arm Image Slice (Ankle 24_1)**

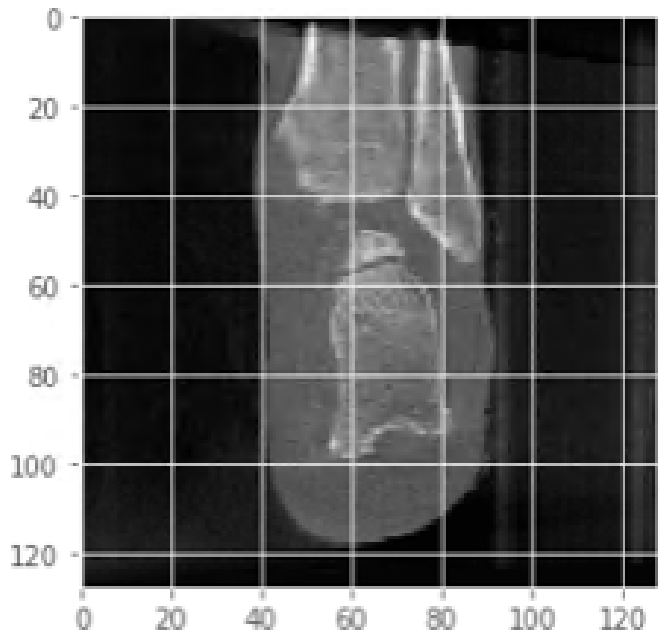

*Figure 25: This is the original slice in sample 24_1 (same slice as last two images).*

**3D Mesh of U-Net Segmented Fibula (Ankle 24_1)**

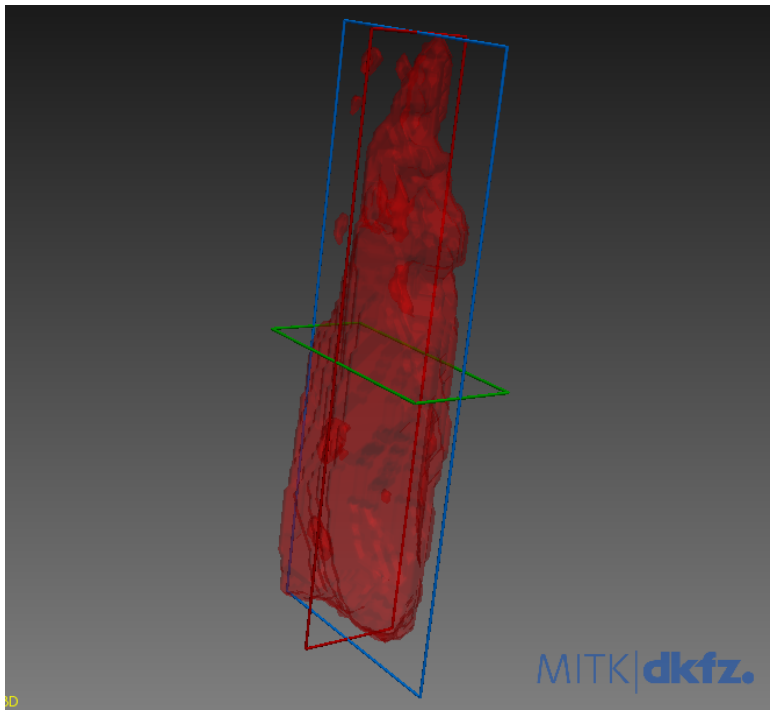

*Figure 26: This is a mesh created from the predicted output from the U-net model for sample 24_1.*

**3D Mesh of Manually Segmented Fibula (Ankle 24_1)**

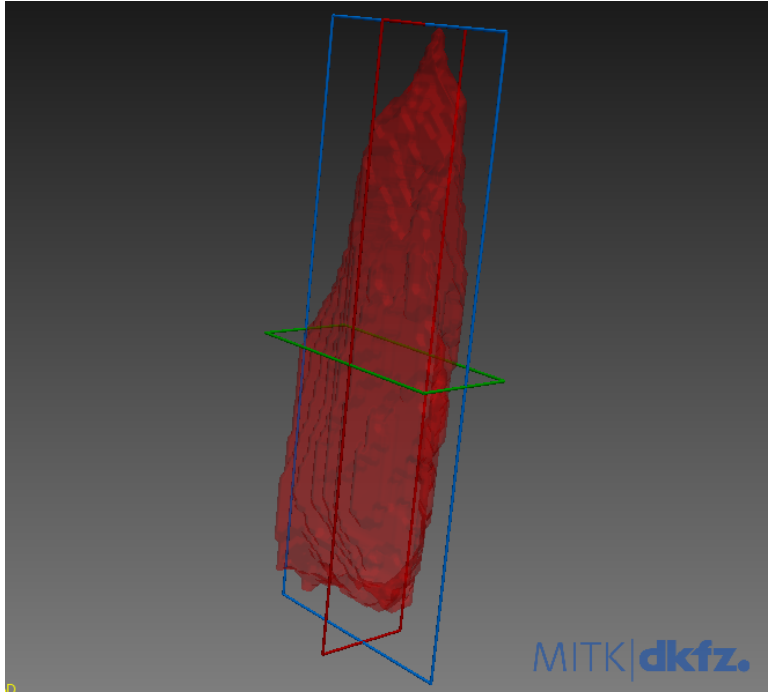

*Figure 27: This is the ground truth mesh for sample 24_1 that was obtained through our manual segmentation.*

As seen above, visually inspecting the meshes reveals that they are very similar. Furthermore, inspecting the meshes on MITK revealed that they were in the same coordinate frame. There are some obvious holes and outliers, but apart from that, the general shapes match. Once we adjust the model so that these visual discrepancies cannot be seen anymore, we will utilize software to calculate error using a closest point to a mesh algorithm. The reason we are not doing this right away is because we must first address the blatant errors in the model. Then we can work on the more specific problems by obtaining error metrics.

## 4. Significance

Initial success of the neural network for segmenting the fibula in C-arm images shows that there is a lot of potential for this approach. This is because the results show that there is promise for not only segmenting other bones in the ankle but segmenting unhealthy ankles could be possible too. Furthermore, after the application of a metal reduction algorithm, it could be possible to to segment metal artifact ankles as well. The significance of this is that in the future it would be possible to accurately segment both ankles of the patient (healthy and unhealthy). This will allow us to eventually solve the problem by flipping and registering the ankles onto each other and then sending instructions to a robot (as explained earlier).

Of course, the future of this project is unknown and is dependent on how our models work. However, the initial results give us confidence to continue pursuing this. Furthermore, there are other models that can also be tried. The cASM model is relatively close to working on C-arm images. So it could be possible that the cASM model by itself or the combinations of the deep learning and cASM models could yield higher segmentation accuracies for some of the bones than simply the deep learning model by itself. There is potential to solve the overall problem, through our current deep learning model or through the other models we have been implementing. It will require a lot more work than what has been done to this point but there is a lot of potential.

We have learned some important skills and concepts as the project has progressed. We learned that cASM is much more computationally intensive than we anticipated. It is also very sensitive to parameter choices. Also, when applying it to C-arm images, instead of CBCT, re-calibration needs to occur for the initial alignment. One of the biggest things we learned with respect to the deep learning approach was how important data pre-processing is in training a neural network. Adequate pre-processing allowed us to obtain results because it reduced the computational burden (which was a lot more than we anticipated) and we were able to get rid of data that would have trained the network incorrectly. We also saw that it required a lot of trial and error to adjust the parameters and end up at the best model possible. We will be applying what we have learned into the rest of the project, hopefully resulting in not only better models but an increase in the speed of model development.

## 5. Management Summary

**Deliverables**

- Minimum: (Expected by 3/11): COMPLETED
    - Working cASM on high quality (CBCT) images
        - Our team will first segment healthy ankle CBCT images without metal artifacts. After consulting with mentors, our segmentation error for each bone should not exceed a mean of 5mm.
        - The purpose of this deliverable is to ensure that our cASM model is working correctly, we will then try to apply it to C-arm images.
        - We will then see if we can use this as a layer in our deep learning network
    - Perform error analysis and show failure modes of ASM model
        - Our mentors do not expect the ASM model to perform adequately for C-arm images with metal artifacts. As a result, our team will create an analysis that documents why the ASM model cannot be implemented on its own.
- Expected: (Expected by 4/11): IN PROGRESS
    - Working Deep Learning Segmentation Model
        - Our team will first segment healthy ankle C-arm images without metal artifacts. After consulting with mentors, our segmentation error for each bone should improve to a mean of 2mm.

- Maximum: (Expected 5/6)
  - Incorporate ankles with metal artifacts into deep learning - cASM model with average segmentation error of 2mm
    - We will apply a metal reduction algorithm (Polyenergetic KCR algorithm) to the Dicom ankle images with metal artifacts and adjust the model in order to incorporate this data
  - Written Paper

**Timeline**

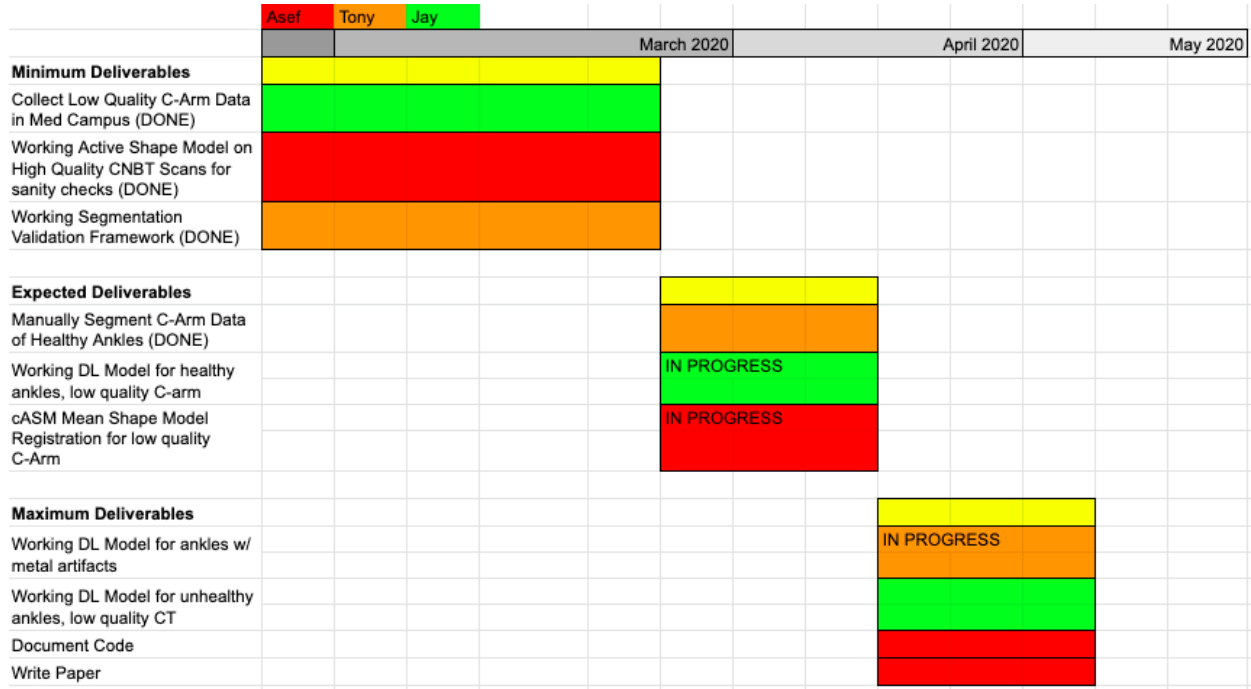

# 6. Acknowledgements

This project is supported by Dr. Jeff Siewerdsen and Wojtek Zbijewski of the JHMI I-STAR Lab. We would like to thank them for their continued support and enthusiasm.

## 8. Technical Appendices

Dr. Wojtek Zbijewski and Dr. Jeff Siewerdsen should have access to the code in the GitHub repository and the C-Arm data

Code:
https://github.com/jmandavilli/CIS2

C-Arm Data:
https://livejohnshopkins-my.sharepoint.com/:f:/r/personal/awu42_jh_edu/Documents/C-Arm?csf=1&web=1&e=03fQIi